# Microrheological Studies of Regenerated Silk Fibroin Solution by Video Microscopy

Raghu A.,[1] Somashekar R.,[2] Sharath Ananthamurthy[1*]

[1]*Department of Physics, Bangalore University, Jnanabharati, Bangalore 560056, India*
[2]*Department of Studies in Physics, University of Mysore, Manasagangotri, India*
*E-mail: asharath@gmail.com

**Abstract**

We have carried out studies on the rheological properties of regenerated silk fibroin (RSF) solution using video microscopy. The degummed silk from the *Bombyx mori* silkworm was used to prepare RSF solution by dissolving it in calcium nitrate tetrahydrate-methanol solvent. Measurements were carried out by tracking the position of an embedded micron-sized polystyrene bead within the RSF solution through video imaging. The time dependent mean squared displacement (MSD) of the bead in solution and hence the complex shear modulus of this solution was calculated from the bead's position information. An optical tweezer was used to transport and locate the bead at any desired site within the micro-volume of the sample, to facilitate the subsequent free-bead video analysis. We present here the results of rheological measurements of the silk polymer network in solution over a frequency range, whose upper limit is the frame capture rate of our camera at full resolution. By examining the distribution of MSD of beads at different locations within the sample volume, we demonstrate that this probe technique enables us to detect local inhomogeneities at micrometre length scales, not detectable either by a rheometer or from diffusing wave spectroscopy.

## INTRODUCTION

Many groups have studied the diffusion and systematic drifts of embedded beads in viscoelastic materials in order to gain insights into mechanisms of various physical processes that occur in these materials. It has been shown that the biological functions of viscoelastic bio-materials can be closely related to their mechanical properties[1,2]. Rheological measurements at micro and nanometre scales can yield information about structure and interaction among the constituent molecules of the material under investigation[3]. Several microrheological techniques have been developed in recent years to study viscoelastic properties of polymer solutions[3]. Video microscopic techniques allow sensitive measurements of the rheological properties of viscous and viscoelastic materials, at length scales, of micro and even nanometres[4,5]. These studies normally involve tracking the position of an embedded particle in a given medium. Initially, these techniques were employed in tracking single particles[1,4], but later, were extended to multiple particle tracking[5]. This improvement has led researchers to extend the use of video microscopy as a rheological tool in the study of viscoelastic materials, as in, studies of the heterogeneities in solutions of actin filaments as a function of concentration[5], studies of F-actin solutions providing new insights into the length scaling behavior of the rheology[6], and gelation processes in block copolypeptide solutions[7], to name a few.

Regenerated silk fibroin (RSF) solution is a biopolymer viscoelastic material (protein) and the silk fibroin molecule is a regular block copolymer, in which ten highly hydrophobic blocks are separated by nine less hydrophobic blocks, making it behave as an amphiphilic block copolymer[8,9]. The unique mechanical properties of the RSF make it especially interesting for many different applications, apart from its wide use in the textile industry[8,10]. It also finds uses in making environmentally sensitive hydrogels, as a chemical valve material, as a food additive, in the cosmetic industry, and as enzyme immobilizers in biosensors[8,10].

Silk fibres from *Bombyx mori* silk worm are produced by a complicated spinning mechanism from its body[2,11]. In studies investigating the macro rheological structure of RSF solution, Xin Chen et al. have investigated using rheometry, an appropriate solvent for *Bombyx Mori* silk by comparing the rheological behavior of the same, in different solvents[8]. Hossain et al. have shown using a viscometer and light-scattering techniques, that aggregation behavior of dilute RSF is different from that of native silk[11]. Ochi et al. have found that divalent metallic ions have a profound influence on the rheology of native silk protein solutions[12]. A study of pH-induced effects on the rheology of native *Bombyx* silk dope has shown that lowering pH (acidification) induces the formation of a gel, detected by a rise in storage modulus[13]. This provides an insight into the possible mechanisms for gel formation in silk fibroin solutions[13]. A recent study on the effect of shear field (spinning rate) and temperature on rheological properties of native silkworm and native spider dope has shown that both dopes behave like polymer melts[2]. These findings provide a better understanding of silk formation from the silkworm, and have tremendous potential in enabling industrial production of silk that matches the qualities of natural silk.

In this work we report measurements of the microrheological properties of RSF in solution, using video microscopy. The position of an embedded tracer polystyrene bead is recorded in silk fibroin solution of different concentrations, in order to measure the storage and loss moduli, and thereby, characterize the viscoelastic properties of this medium. While the upper value of frequency of the measurements reported here are limited by the capture rate of our CMOS camera, to 15 Hz, the method of directly tracking the positions of an ensemble of beads in the medium enables us to characterize the local variations in the viscoelasticity of silk polymer fibroin. In addition, we are able to identify the inhomogeneity of the medium at higher concentrations, by monitoring the mean squared displacement of beads in different positions within the sample. These measurements yield different values of the shear moduli within the same sample and thereby reflect the local inhomogeneities. Such features are averaged out and therefore, not detectable, when measurements are carried out using a rheometer or by diffusing wave spectroscopy.

**EXPERIMENTAL SECTION**

**Sample Preparation**

Raw *Bombyx mori* silk primarily consists of two components, hydrophobic silk fibroin and hydrophilic sericin. In this study we were interested in the rheological behavior of the fibroin molecules, and therefore we have used air dried, degummed, silk fibres from *Bombyx mori*. A 75:25 (weight ratio) mixture of Ca $(NO_3)_2$ $4H_2O$ and absolute methanol ($CH_3OH$) is prepared and used as a solvent[8,13]. Although other solvents may also be used for dissolving *Bombyx mori* silk[8], we chose this solvent as it has the strongest dissolving capacity for the silk fibroin[8]. A known weight of clean degummed *Bombyx mori* silk is added to a known volume of solvent and the mixture, stirred at 75°C for 10 minutes to dissolve the fibre. This cleaned undialyzed regenerated silk fibroin (RSF) solution is used for our studies reported here. Undialyzed salt solution of silk fibroin was chosen for the work, since the presence of metallic ions ($Ca^{2+}$) in solution helps to maintain the fibroin's structure in solution[12] and moreover the undialyzed solution contains oriented silk fibres as compared to the dialyzed solution, wherein the fibroin is in a predominantly aqueous environment[14]. Three different concentrations, 1.00%, 0.50% and 0.15% (w/v) of RSF were prepared and used for microrheological analysis. The respective concentrations of the solutions in g/mL are $10 \times 10^{-3}$, $5 \times 10^{-3}$ and $1.5 \times 10^{-3}$. In this work we will follow the convention of stating concentrations in the former way.

Polystyrene beads (Cat.no. 07310-15, 0.989 ± 0.02 μm, Polysciences Inc., USA) in aqueous solution are used as probes to measure the relation between stress and deformation in the viscoelastic materials. They are added to the samples in low concentrations (1μL in 1mL of RSF) to avoid inter-bead interactions. The sample is agitated by shaking gently to achieve uniform mixing of beads in solution for about one minute and then allowed to stabilize for 15 minutes before conducting trials. About 70μL of sample was used for each trial, and the sample was poured carefully into a single cavity microscope slide and covered with number 1 grade cover slip to avoid air bubbles in the sample. Molten wax was used to seal the slide. The chemicals and reagents used in this work were HPLC grade supplied by Merck Chemicals, India.

**Video Microscopy**

Video microscopy makes use of processing time-sequenced image frames of the embedded beads in a given sample in order to study the rheological properties of that sample. In our open microscope (home built) setup[15], the sample is visualized on a computer monitor through a CMOS camera (EDC-3000, Electrim Corporation, USA), after magnifying the sample image though a 100× oil immersion objective (1.25 NA, EA100× Long Barrel, Olympus, Japan). The spatial (temporal) resolution of our imaging system is 63nm/pixel (15 fps) which is determined by camera pixel density and optical magnification[4,6]. Movies of Brownian motion of the beads in a given medium are stored on to a computer hard disk (Intel Pentium IV, 1GB RAM, 80 GB HDD) for further processing by use of image analysis algorithms[4]. We have used a feature finding algorithm, originally by J.C. Crocker and D.G. Grier[4], later incorporated into IDLVM (Interactive Data Language Virtual Machine, RSI Systems, USA) software by Ryan Smith and Gabe Spaulding[16]. This program can detect displacements of a bead with a precision of a few tenths of a nanometre[4,7] in sequenced image frames. The final resolution achievable depends not only on the choice of algorithm but also on signal to noise ratio of the raw data. Nearly 1000 images are acquired for each measurement, after focusing the microscope well above (about 30 μm) the cover slip wall, to avoid monitoring any beads interacting with the cover slip. When necessary, an optical tweezer[15] is used to transport and locate the beads at a desired site. Through multi particle tracking 3-5 beads are tracked simultaneously for each field of view, and this helps save time while obtaining good statistics of the data from only a few movies.

**Method**

Using the time dependent position information of the thermally driven bead of radius *a* embedded in the given medium at a temperature *T*, we can calculate its time dependent mean squared displacement (MSD=$\langle \Delta r^2(\tau) \rangle$) in that medium, defined as[3,17],

$$\langle \Delta r^2(\tau) \rangle = \langle |r(t+\tau) - r(t)|^2 \rangle \quad (1)$$

where '$\tau$' is the lag time and $\langle \rangle$ denotes the average being calculated for all starting times '*t*'.

Custom programs are written in LabVIEW (National Instruments, USA) platform to obtain MSD of a bead from the raw data using equation (1). For a viscoelastic medium, this MSD can be related to the complex shear modulus $G^*(s)$ in Laplace domain through the generalized Stokes-Einstein relation (GSER) given by[17],

$$\langle \Delta r^2(s) \rangle = \frac{k_B T}{\pi a s G^*(s)} \quad (2)$$

where '$s$' is the Laplace frequency, and '$k_B$' is the Boltzmann constant.

The analytic continuation method developed by Mason[18] can be used to obtain the complex shear modulus $G^*(\omega)$, storage modulus $G'(\omega)$ and loss modulus $G''(\omega)$, in Fourier domain from MSD values. Dasgupta et al.[19] have empirically modified these equations to improve the accuracy of moduli determined from the data where MSD data displays greater curvature. These expressions are,

$$G'(\omega) = G^*(\omega) \{1/[1+\beta'(\omega)]\}$$
$$\cos\left[\frac{\pi \alpha'(\omega)}{2} - \beta'(\omega)\alpha'(\omega)\left(\frac{\pi}{2}-1\right)\right] \quad (3)$$
$$G''(\omega) = G^*(\omega) \{1/[1+\beta'(\omega)]\}$$
$$\sin\left[\frac{\pi \alpha'(\omega)}{2} - \beta'(\omega)[1-\alpha'(\omega)]\left(\frac{\pi}{2}-1\right)\right]$$

Where,

$$G^*(\omega) = \frac{k_B T}{\pi a \langle \Delta r^2(1/\omega) \rangle \Gamma[1+\alpha(\omega)](1+\beta(\omega)/2)} \quad (4)$$

Here $\alpha(\omega)$ and $\beta(\omega)$ are the first and second order logarithmic time derivatives of the MSD, $\alpha'(\omega)$ and $\beta'(\omega)$ are the first and second order logarithmic frequency derivatives of $G^*(\omega)$ and are obtained by fitting the data locally to a second order polynomial. This method has good accuracy with less than 5% error for the dominant modulus of the two in the whole frequency range[19].

## RESULTS

### Viscous fluids

Analysis of the Brownian motion of free beads in water using the image processing code[16] reveals that we can detect displacement of bead positions as small as 4nm. This sets the resolution limit of the data gathered with our set up. We have analyzed the position information of beads stuck to the cover slip, for nearly 900 frames. An average standard deviation of 33 nm in the stuck bead position displacement was observed from an ideal value of zero, for stationary stuck beads. This noise is a reflection of the artefacts inherent in the imaging system of our system thereby setting the upper limit on the accuracy of the data in our setup.

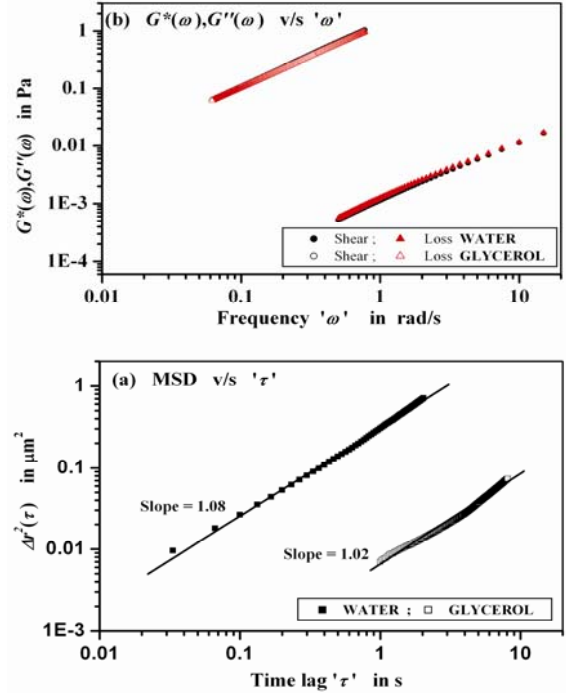

Figure 1 : (a) MSD $\langle \Delta r^2(\tau) \rangle$ of water (■) and glycerol (□) as a function of lag time. (b) Shear moduli $G^*(\omega)$ of water (●) and glycerol (○) coinciding with corresponding loss moduli $G''(\omega)$ of water (▲) and glycerol (△) at all measured frequencies.

To calibrate our set up, we have used 0.989 μm polystyrene beads in pure viscous media such as double distilled water and pure glycerol. From the bead position data, we have calculated the MSD of beads in both solutions using equation (1). MSD of 14-16 beads were measured at different parts of each sample and used them to calculate an ensemble averaged MSD. Figure 1(a) shows the measured MSD as a function of lag time, of water and glycerol. The slope of the curves of Figure 1(a) confirms that the media is pure viscous[5,7]. These MSD values are used to calculate the shear moduli and loss moduli by (3) & (4) and are shown in Figure 1(b), for the two fluids. Note that, we have omitted the data of MSD below 0.040 μm$^2$, the operational limit of our setup. Omission of this data limits the upper frequency value of the measured rheological parameters. The storage moduli are zero at all frequencies as is expected for pure viscous materials. The loss modulus grows linearly with frequency obeying the relation[18],

$$G''(\omega) = \eta \omega \quad (5)$$

The viscosity '$\eta$' of the medium can be calculated by the slope of Figure 1(b), by fitting log $G''(\omega)$ to $\log(\omega) + \log(\eta)$. This fit yields the viscosity of water and glycerol as 1.08±0.04 mPas and 0.94±0.06 Pas respectively, and these are in agreement with the standard values at 22°C. These measurements serve to calibrate our setup.

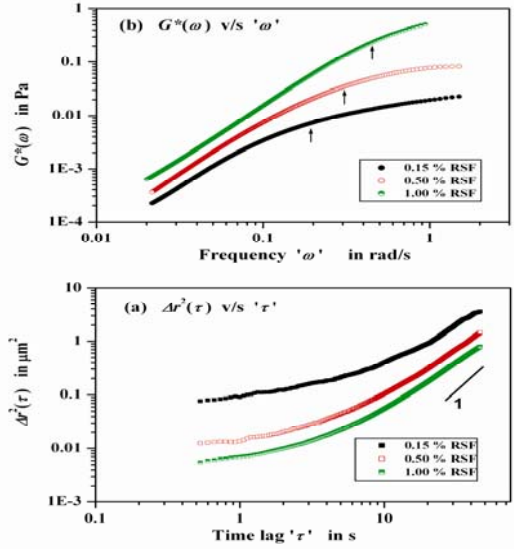

**Figure 2:** (a) MSD $\langle \Delta r^2(\tau) \rangle$ of three RSF concentrations (■ 0.15 % wt; □ 0.50 % wt; ▬ 1.00 % wt). A line with slope one at higher lag times is drawn for comparison. (b) Corresponding shear moduli $G^*(\omega)$ (● 0.15 % wt; ○ 0.50 % wt; ◐ 1.00 % wt) of three RSF concentrations.

### RSF Solution

We have used the calibrated set up to study RSF solution. The ensemble averaged MSD of 15-18 beads in each concentration of solution is obtained by the method explained earlier. Figure 2(a) shows the MSD behavior of beads in RSF, for three concentrations, against the lag time '$\tau$'. Each curve in this plot has different slopes (between 1 and 0) at different regions of lag time, indicating that silk fibre solution exhibits both elastic and viscous nature[7].

Figure 2(b) shows the frequency dependence of the shear moduli for different concentration of RSF solution. The magnitude of shear modulus increases with increase in the concentration of RSF solution. According to the elementary unit model of Ochi et al.[11], with increase in RSF concentration, the number of fibroin junctions increases due to overlapping of the fibroin molecules and this in turn results in an increase in the shear modulus of the sample. This is consistent with our observations. In the lower frequency region (< 0.1 rad/s) the shear modulus increases with frequency almost linearly for all the three samples, but at higher frequencies shows less increment with frequency, suggesting the dominance of elastic component over viscous component of the sample, as identified by the change in slope[7]. The transition from liquid-like to solid-like behavior is a function of the RSF concentration in solution and the frequency at which this occur increases with concentration, as indicated by the arrow marks in the plot. The solid-like behaviour of the RSF at higher frequencies suggest that, on short time scales the beads appear to be held rigidly in the matrix of silk fibroin network as compared to longer time scales.

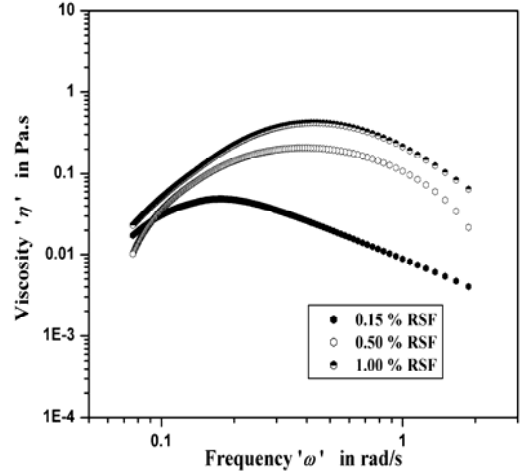

**Figure 3:** Variation of viscosity $\eta$ with frequency for three RSF concentrations (● 0.15 % wt; ○ 0.50 % wt; ◐ 1.00 % wt).

Viscosity (Fig 3) for the three RSF concentrations obtained using equation (5) shows a small initial increase and then a decrease at higher frequencies, which can be attributed to shear thickening and shear thinning of the sample respectively. This trend is similar to that observed in rheometric studies[8]. The differences in the measured values are attributable to the fact that rheometry involves the application of an external stress/strain field to the sample where as video microscopy is a passive technique wherein the strain field is produced by the thermally diffusing beads. A second reason could be the electrostatic interaction between the polystyrene beads and the amphiphilic fibroin molecules causing slightly enhanced values of shear thickening and shear thinning from those measured using a rheometer. Upper frequency measurement restrictions preclude us from comparing our data with that acquired using a rheometer[8] to frequencies above 2.25 rad/s.

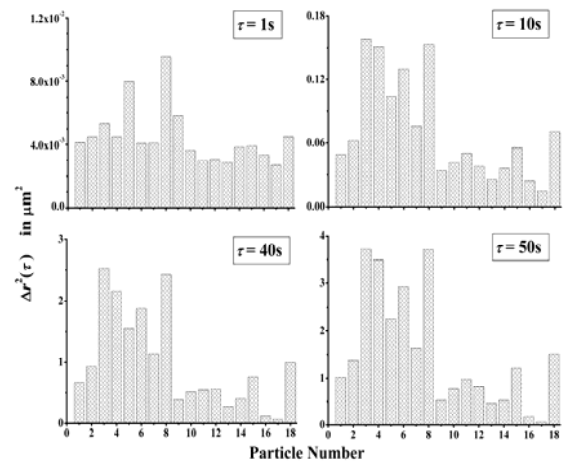

**Figure 4:** Bar plot of MSD of beads in 1.0% RSF at different lag times. At 50 s the MSD distributions showing two bead families, one (family 1) with MSD less than 1 μm² and another (family 2) with MSD more than 1 μm².

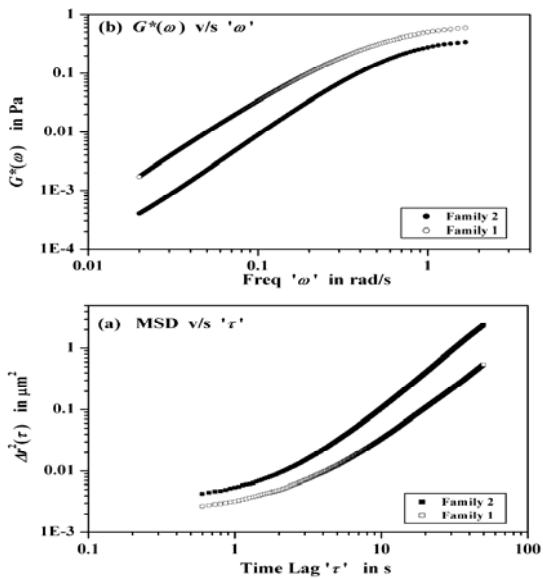

**Figure 5** : (a) MSD of bead family 1 (□) is less than that of family 2 (■) at all frequencies suggesting latter is relatively free compared to the former. (b) Shear moduli of the bead family 1 (○) and bead family 2 (●).

Next, we examine the distribution in MSD of a set of beads in 1.0% RSF solution at different lag times. Figure 4 shows MSD values of 18 beads at lag times of 1 s, 10 s, 40 s and 50 s. At small lag times the MSD distribution is nearly uniform. As the lag time increases the MSD distribution starts to show nonuniformity. At a lag time of 10 s and above, two distinct families of MSD values appear. At lag times of 50 s one can clearly distinguish the family of beads showing MSD less than 1 μm$^2$ and the other family with MSD more than 1 μm$^2$. This serves to display the structural heterogeneity in the RSF solution. Apgar et al.[5] have identified the presence of inhomogeneity in F actin network as defects due to the presence of small and large pores within the sample. Similarly we expect reason for the structural heterogeneity in RSF was due to the presence of pores of different size within the sample. When the MSD values of the two families of beads are analysed separately, this results in differing values of shear moduli, as shown in Figure 5(b). The lower MSD values (Figure 5 (a)) of the beads are attributed to the presence of cages, due to small pore size, inside the sample wherein the beads get trapped and thereby, show restricted diffusion. This corresponds to regions of greater stiffness of the polymer network in the sample. Other beads showing larger values of MSD correspond to relatively free diffusion within the sample.

## CONCLUSION

Our measurements demonstrate that RSF at various concentrations displays viscoelastic properties. The transition from liquid-like to solid-like behavior is a function of the RSF concentration in solution and the frequency at which this occur increases with concentration. Through video microscopic technique we have established, qualitatively, that the RSF polymer solution displays structural heterogeneity, at the micron length scales. Further studies, in progress, are necessary to quantify the heterogeneity of RSF solution. We are, further, studying RSF rheology in various physical and chemical conditions to examine the concentration dependence of the scaling laws for amphiphilic block co-polymers such as RSF[1,9].

## ACKNOWLEDGEMENTS

The authors acknowledge a research grant from the Department of Science and Technology, Govt. of India (under the Nano Science and Technology Initiative, Phase II) to carry out this work. They thank Dr. R. Damle, Department of Physics, Bangalore University for loan of the magnetic stirrer used in the synthesis of the RSF.